\documentclass[12pt,fullpage,epsfig]{article}
\def\beq{\begin{equation}}
\def\eeq{\end{equation}}
 
\begin{document}
\begin{center}
{\Large{\bf   Steady State and Relaxation Spectrum of  the Oslo Rice-pile 
Model}}\\[2cm] 
 
{\large{\bf Deepak Dhar}}\\
Department of Theoretical Physics, \\
Tata Institute of Fundamental Research, \\
Homi Bhabha Road, Mumbai 400~005, INDIA\\ [2cm]
 
\end{center}
\bigskip
\begin{abstract} 

      We show that the one-dimensional Oslo rice-pile model is a special
case of the abelian distributed processors model. The exact steady state
of the model is determined. We show that the time evolution operator
${\cal W}$ for the system satisfies the equation ${\cal W}^{n+1} = {\cal
W}^{n}$ where $n = L(L+1)/2$ for a pile with $L$ sites.  This is used to
prove that ${\cal W}$ has only one eigenvalue $1$ corresponding to the
steady state, and all other eigenvalues are exactly zero. Also, all
connected time-dependent correlation functions in the steady state of the
pile are exactly zero for time difference greater that $n$. Generalization
to other abelian critical height models where the critical thresholds are
randomly reset after each toppling is briefly discussed.

\end{abstract}

\section{Introduction}

In their pioneering work in 1987,  Bak et al proposed the well-known
sandpile model as a paradigm for self-organized criticality \cite{BTW}.
Since then, many different variants of sandpiles have been studied. These
include models where the height of the pile is a real variable
\cite{zhang}, different toppling rules \cite{kadanoff}, with preferred
direction \cite{dr}, stochastic topplings \cite{manna}, models with fixed
energy \cite{fixedenergy}, etc.. While there is a fair amount of
understanding by now about many of these models, the understanding of
different universality classes of critical behavior possible, and of
relationship to the critical steady states seen in other driven systems
remains unsatisfactory \cite{universality}.

   The static properties of the BTW sandpile model can be related to the
equilibrium properties of $q \rightarrow 0$ limit of the $q$-state Potts
model and the statistics of spanning trees \cite{majumdar}. The directed
model \cite{dr} can be related to the well-known voter and the Takayasu
aggregation models \cite{eplf}. However, the exponents characterizing the
avalanche distributions in sandpile models are not determined completely
by the critical exponents of the $q=0$ Potts model. We have argued that
the generic behavior of sandpile like models is in the universality class
of directed percolation \cite{mohanty}. However, the critical behavior of
special models, like the BTW or the Manna models, is certainly not in this
class.  Paczuski and Boettcher have given arguments relating the exponents
of sandpile model to that of growing interfaces \cite{interfaces}.  
Generically, the growing surfaces are expected to be in the universality
class of Kardar-Parisi-Zhang type models, not  of directed percolation. It
seems desirable to have a closer look at the relation between sandpile
models and growing surfaces. This relationship is most clearly seen
\cite{pruessner} in the simple Oslo rice-pile model \cite{frette1}.

  The Oslo rice-pile is one of the simplest of models of self-organized
criticality and seems to be able to describe quite well the real
experiments on rice-piles \cite{frette2}.  It shows non-trivial avalanche
exponents even in one dimension.  There are some exact results known about
this model \cite{oslo1}, and the values of avalanche exponents are known
quite accurately \cite{oslo2}. The damage propagation in the model has
also been studied \cite{oslo3}.  It seems worthwhile to try to see if the
model can be solved exactly. While attempts in this direction have not
been successful so far, it does seem to have a rather unusual mathematical
structure.  In this paper, I will try to outline some of these properties,
and hope that this will encourage further work.

\section{Definition of the Model}

The Oslo rice-pile model is a one-dimensional cellular automaton model
with stochastic toppling rules.  It is defined as follows \cite{frette1}:
We consider a line of $L$ sites, labelled by integers $1$ to $L$. At each
site $i$, there is a non-negative integer height variable $h_i$, called
the height of the pile at that site. We start with a configuration in
which all heights are zero. At each site $i$, there is also a variable
$\sigma_i^c$, called the critical threshold, which takes values $2$ or
$1$ with probabilities $p$ and $q = (1-p)$ respectively, independently at
each site.

We shall call $h_i - h_{i+1}$ as the local slope of the pile at site $i$,
and denote it by $\sigma_i$. We define $h_{L+1} = 0$, so that the local
slope at the site $i=L$ is equal to $h_L$. We call a site $i$ stable if
$\sigma_i \leq \sigma_i^c$. A configuration is said to be stable, if all
sites in the configuration are stable.

If a site $i$ is unstable, it relaxes by sending one grain to site on its
right. In this process, $h_i$ will decrease by $1$, and $h_{i+1}$ will
increase by $1$. If a configuration has more than one unstable sites, then
they are updated in parallel. ( We will show later that actually the order
of topplings does not matter).  After each toppling at a site, the
critical threshold at that site is reset, and takes a new random value $1$
or $2$ with probabilities $q$ and $p$ respectively.

The system is driven by adding a grain at the leftmost site $i=1$. This
increases the local height by $1$. If this leads to an unstable
configuration, it is relaxed by topplings till a stable configuration is
reached. And then we again add a grain at $i=1$.

If we evolve the system like this, it reaches a self-organized critical
state at large times, where the distribution of avalanches shows a power
law tail with an L-dependent cutoff.

\section{The Abelian Property}

It is convenient to specify the configurations in terms of local slopes
$\sigma_i$. A particular configuration $\{h_i\}$ corresponds to a unique
set of slope variables $\{\sigma_i\}$, and vice versa. In terms of the
slope variables, the evolution rules of the rice-pile become the 
following:\\
i) If a particle is added from the left, $\sigma_1$ increases by $1$.\\
ii)a. If $\sigma_i > \sigma_i^c$ with $2 \leq i \leq L-1$, then 
$\sigma_i$
decreases by $2$, and $\sigma_{i-1}$ and $\sigma_{i+1}$ increase by $1$.\\
\hspace*{3 mm} b. If $\sigma_1 > \sigma_1^c$, then, $\sigma_1$ decreases by 
$2$, and 
$\sigma_2$ increases by $1$.\\
\hspace*{3 mm} c.If $\sigma_L > \sigma_L^c$, then, $\sigma_L$ decreases by 
$1$,
$\sigma_{L-1}$ increases by $1$.\\
iii) After a toppling at  site $i$,  $\sigma_i^c$ is reset to a new 
randomly chosen value.

It is interesting that these rules are same as for a one-dimensional
critical height model, where the critical threshold is a random variable,
and its value at a site is reset randomly after each toppling at that
site. This correspondence to an {\it abelian} critical height model is not
possible in higher dimensions, or if grains in the Oslo model are added at
all sites, and not only at the left end.

It is convenient to work with this one-dimensional critical height model,
in which the ``particles" are the ``slopes variables" of the original
model. To avoid confusion, in this paper we shall call the rice-particles
of the original model ``grains", and use the word ``particles" to refer to
the slope variables. Thus, in a single toppling event at site $i$, $( 2
\leq i \leq L-1 $), a grain is transferred from $i$ to $(i+1)$, but two
particles are moved from $i$, one each to the sites $(i-1)$ and $(i+1)$.
The topplings at the boundary sites can be similarly described.

As another simplification, we need not keep track of the critical
thresholds, if we adopt the following rule for relaxation {\it whenever a
new particle is added to a site} : If, after addition, the new height is
$> 2$, the site topples;  if the height is $2$, it topples only with
probability $(1-p)$; and if the new height is $1$, it does not topple.
Clearly, the actual evolution is same as under the original definition.

This model, then is a special case of the general Abelian distributed
processors model defined earlier \cite{eplf}. We consider each site as a
finite-state automaton, with a local pseudo random number generator
(PRNG), and a single integer giving the number of particles at the site.
If an added particle makes the number of particles greater than the
current threshold, a fixed number of particles, as specified by the matrix
$\Delta$, are transferred to other sites. A new random number is drawn
from the PRNG, and used to choose a new threshold. It is easily shown that
in any configuration with more than one unstable sites, the final state of
this system does not depend on the order in which these sites are relaxed,
for any particular sequence of thresholds generated by the PRNG. This 
establishes the abelian property  for the rice-pile model.

\section{Operator Algebra}

Let us denote by ${\cal V}$ a linear vector space spanned by its basis
vectors $|{\cal C}\rangle$, where $|{\cal C}\rangle$ are the stable configurations of
the rice-pile. We shall characterize the configurations ${\cal C}$ by the
slope variables $\{\sigma_i\}$.  There are $3^L$ such configurations for a
chain of length $L$. Note that the values of the critical slopes
$\sigma_i^c$ are not specified in ${\cal C}$.

The state of the rice-pile at time $t$ is characterized by a probability
vector $|Prob(t)\rangle$, which is an element of ${\cal V}$. If the probability
that the rice-pile occurs in configuration ${\cal C}$ at time $t$ is given
by $p({\cal C},t)$, we write
\begin{equation}
|Prob(t)\rangle =  \sum_{\cal C} p({\cal C},t)|{\cal C}\rangle.
\end{equation}

We now define linear operators $a_i$ ( $i = 1$ to $L$ ) acting on ${\cal
V}$, by their actions on the basis vectors $|{\cal C}\rangle$. Starting from the
stable configuration ${\cal C}$, we increase the slope variable $\sigma_i$
by $1$, and allow the resulting configuration to relax if necessary. If
the resulting configuration is ${\cal C'}$ with probability $w_i({\cal
C'|C})$, we define
\begin{equation}
a_i |{\cal C}\rangle = \sum_{\cal C'} w_i({\cal C'|C}) |{\cal C'}\rangle.
\end{equation}
We then have
\begin{equation}
[ a_i, a_j ] = 0, {\rm ~~for ~~all~~} i, j.
\end{equation}
The time-evolution of the system is Markovian. Let $|Prob(t)\rangle$ be the
probability vector of the state of the system after $t$ paticles have been
added, and the system is allowed to relax. We then have the master
equation

\begin{equation}
|Prob(t+1)\rangle = {\cal W}~ |Prob(t)\rangle,
\end{equation}
which defines the time evolution operator ${\cal W}$. For the rice-pile
model, with particles added only at the left end, clearly, we have ${\cal
W} =a_1$.

Let us now give a more explicit representation of the operators $\{a_i\}$.
If $a_i$ acts on a configuration where $\sigma_i = 0$, we get a state with
$\sigma_i$ increased to value $1$, and all other $\sigma$'s are unchanged.
\begin{equation}
a_i |\ldots, \sigma_i=0,\ldots\rangle~~ =~~ | \ldots, 
\sigma_i=1,\ldots\rangle.
\end{equation}
If $a_i$ acts on a configuration with $\sigma_i=1$, then with probability
$p$, it increases its value to $2$.  With probability $(1-p)$, it causes a
toppling there, which would change $\sigma_i$ to $0$, and add a particle
at sites $(i \pm 1)$.
\begin{equation}
a_i |\ldots, \sigma_i=1,\ldots\rangle~~ =~~p  | \ldots, \sigma_i=2,\ldots\rangle +
q  a_{i-1} a_{i+1} | \ldots, \sigma_i=0,\ldots\rangle.
\end{equation}
Acting on a configuration with $\sigma_i=2$, $a_i$ will always cause a 
toppling, and we have
\begin{equation}
a_i |\ldots, \sigma_i=2,\ldots\rangle~~ =~~   a_{i-1} a_{i+1} | \ldots, 
\sigma_i=1,\ldots\rangle.
\end{equation}

These equations fully define the action of the operators $a_i$. They also 
hold for the boundary sites $i=1$, and $i=L$, if we assume the conventions
\begin{equation}
a_0 = 1,{\rm ~~and ~~} a_{L+1} = a_L.
\end{equation}

Applying these rules repeatedly, we can determine the effect of any of the
the operators $\{a_i\}$ on any stable configuration. For example, consider
the case $L=2$. Let us determine the result of $a_1$ acting on $|2,2\rangle$.  
We get,
\begin{equation}
a_1 |2,2\rangle ~=~ a_2|1,2\rangle ~=~ a_1 a_2 |1,1\rangle ~=~ 
a_1(~ p |1,2\rangle + q a_1  |1,1\rangle).
\end{equation}
These two terms can be evaluated further as
\begin{equation}
a_1 |1,2\rangle = p|2,2\rangle + q a_2|0,2\rangle, 
\end{equation}
with
\begin{equation}
a_2 |0,2\rangle = a_1 a_2 |0,1\rangle = a_2 |1,1\rangle = 
p |1,2\rangle + q  a_1 |1,1\rangle, 
\end{equation}
and
\begin{equation}
a_1 |1,1\rangle = p |2,1\rangle + q a_2 |0,1\rangle = p |2,1\rangle + 
pq |0,2\rangle + q^2 |1,1\rangle.
\end{equation}
Putting all these together, we get
\begin{equation}
a_1 |2,2\rangle = p^2 |2,2\rangle  + (1 - p^2)~[~  p |1,2\rangle + qp |2,1\rangle + q^2 p | 0,2\rangle + 
q^3 |1,1\rangle]
\end{equation}
We also note that these operators satisfy the equations
\begin{equation}
\label{reduce}
a_i^3 = a_{i-1} a_i a_{i+1}, {\rm ~~ for~~ all~~ } i.
\end{equation}
This is equivalent to the observation that on  adding three particles at 
site 
$i$, a
toppling there must occur at site $i$, so that it is same as adding a 
particle each at
the sites $i-1, i$ and $i+1$.
 
Given any product of the form $a_2 a_1 a_7 a_2 a_2 \ldots$, we first use
the commutativity property to bring it to the form $a_1^{n_1} a_2^{n_2}
\ldots$, where $n_1, n_2, \ldots$ are non-negative integers.  Then, we use
the reduction rules Eq. (\ref{reduce}) to express it as a sum of lower
powers such that $n_i \leq 2$ for all $i$. Consider evaluating $a_1^r$ for
$r = 1,2,\ldots$. As the number of possible answers is finite, we must
have a minimum value of $r$, such that $a_1^r$ is equal to a lower power
of $a_1$. Normally, one would guess that $r$ is of order {\mbox exp(L)}, 
the
number of different terms allowed. Interestingly, it turns out to be much
smaller, and we now show that $r = L(L+1) + 1$ for all $L$.

Using Eq.(\ref{reduce}), it is  straightforward to show that
\begin{equation}
a_1^{L(L+1)} = a_1^2 a_2^2 a_3^2 \ldots a_L^2.
\end{equation}
Multiplying both sides of  this equation  by $a_1$, and again using 
the reduction rules Eq.({\ref{reduce}),  we get
\begin{equation}
a_1^{L(L+1)+1} =  a_1^2 a_2^2 a_3^2 \ldots a_L^2.
\end{equation}
As $a_1 = {\cal W}$, we get
\begin{equation}
\label{wll}
{\cal W}^{L(L+1)+1} = {\cal W}^{L(L+1)}.
\end{equation}

This equation holds as an operator equation, over the $3^L$ dimensional
vector space ${\cal V}$, the space of all stable configurations. This has
the remarkable consequence that eigenvalues of ${\cal W}$ are either $0$
or $1$.  ${\cal W}$ is a stochastic matrix. As there is a finite
probability that the maximal state $|2,2,\ldots,2\rangle$ can be reached from
any stable configuration, all recurrent configurations are reachable from
any other, and there is a unique steady state for this Markov process.  
This implies that there is  only one eigenvector of ${\cal W}$ with
eigenvalue $1$. Let us call it $|\Psi_{st}\rangle$ Thus all the other $3^L -1 $
eigenvalues of ${\cal W}$ are $0$.

From Eq.(\ref{wll}), it follows that for any vector $|\phi\rangle $ in ${\cal
V}$, ${\cal W}^{L(L+1)} |\phi\rangle$ is proportional to $|\Psi_{st}\rangle$. Also, as
${\cal W}$ preserves the probability sum, for any initial basis vector
$|{\cal C}\rangle$, we get
\begin{equation}
\label{weq}
{\cal W}^{L(L+1)} ~ |{\cal C}\rangle = |\Psi_{st}\rangle.
\end{equation}
A simpler way to determine  $|\Psi_{st}\rangle$ is to use the equation
\begin{equation}
|\Psi_{st}\rangle = a_1 |2,2,2\ldots 2\rangle.
\end{equation}

This equation is interesting, as it says that if we take the initial
configuration as the one in which each $\sigma_i$ has the highest allowed
value, and just add one more grain, the stochastic evolution would result
in different final stable configurations with probabilities exactly equal
to their values in the steady state. To prove this result, we need only
note that for any stable configuration,
\begin{equation}
a_i |\dots,\sigma_i=2,\ldots\rangle = a_i^3 
|\ldots,\sigma_i=0,\ldots\rangle.
\end{equation}
Then we can write  
\begin{equation}
a_1|2,2,2\ldots 2\rangle  = a_1^3|0,2,2\ldots\rangle = a_1 a_2|0,2,2\ldots\rangle = a_1 a_2^3|0,0,2\ldots\rangle
\end{equation}
\begin{equation}
 = a_1^2 a_2 a_3|0,0,2\ldots\rangle = a_1^2 a_2 a_3^3 |0,0,0\ldots\rangle 
= etc.,
\end{equation}
finally giving
\begin{equation}
a_1 |2,2,2\ldots\rangle = a_1^2 a_2^2 a_3^2 \ldots a_L^2|0,0,0\ldots\rangle~ 
= ~|\Psi_{st}\rangle.
\end{equation}

In fact, it is possible to get a result stronger than Eq.(\ref{weq}). We
note that most of the stable configurations are transient, and do not
occur in the steady state. The different recurrent configurations of the
Oslo rice-pile have been characterized \cite{oslo1}. For any recurrent
configuration ${\cal C}$ with $m$ grains, arguing as above, we can show
that
\begin{equation}
a_1^{L(L+1)-m} |{\cal C}\rangle = a_1^2 a_2^2 a_3^2\ldots a_L^2|0,0,0\ldots0\rangle
\end{equation}
Since the lowest allowed value of $m$ in recurrent configurations is 
$L(L+1)/2$, corresponding to the 
configuration $|1,1,1,\dots\rangle$, we get that for any recurrent configuration 
${\cal R}$,
\begin{equation}
\label{corr}
a_1^{L(L+1)/2 } |{\cal R}\rangle = |\Psi_{st}\rangle
\end{equation}

Let $X(t)$ be some scalar observable whose value depends on the stable
configuration of the pile after $t$ grains have been added. Then any
particular evolution of the rice-pile will generate a stochastic
time-series $\{X(j)\}, j = 1,2, \ldots $. For example, $X(t)$ may be the 
total height of the
pile $h_1$ at time $t$, or the number of grains in the pile. Clearly, this
time-series has nontrivial correlations. For example, $h_1(t+1) \leq
h_1(t)+1$. For any two such observables $X(t)$ and $Y(t)$, we define the
connected time-dependent correlation function $C_{XY}(\tau)$ in the steady
state as
\begin{equation}
C_{XY}(\tau) = \langle X(t) Y(t + \tau)\rangle -\langle X(t)\rangle 
\langle Y(t +\tau)\rangle
\end{equation}
However, from Eq.(\ref{corr}), it follows that the conditional expectation
value of $Y(t + \tau)$, given that the observable $X$ has a particular
value $X(t)$, must be equal to its unconditional expectation value,
whenever $\tau \geq L(L+1)/2$.  Thus, $C_{XY}(\tau)$ is exactly equal to
zero for $\tau \geq L(L+1)/2 $, for all observables $X$ and $Y$.

For $\tau < L(L+1)/2$, the correlation function $C_{XY}(\tau)$ is not 
always zero. This correlation function can be nontrivial, 
even though ${\cal W}$ has only one nonzero eigenvalue, because ${\cal W}$ 
is non-hermitian. This is easily checked for the simple cases $L= 2, 3$. 

\section{Generalizations}

It is straightforward to generalize the discussion given above to more
general abelian sandpile model (ASM) \cite{dd90}. Consider a stochastic
sandpile model defined on a set on $N$ vertices labelled by integers $1$
to $N$. At each site is a height variable $z_i$, which takes non-negative
integer values. There is an $N \times N$ integer matrix $\Delta$, which
specifies how particles are transferred under topplings: If $z_i$ is
greater than or equal to a site-dependent threshold value $z_{i,c}$, there
is a toppling at site $i$, the heights at all sites $j$ are updated
according to the rule

\begin{equation}
z_j \rightarrow z_j - \Delta_{i,j}.
\end{equation}
The matrix $\Delta$ is assumed to satisfy the good-behavior conditions, 
same as for ASM \cite{dd90}.
After each toppling at a site $i$, the critical threshold at that site is
randomly reset to a new value $z_{ic} = r$, with $r$ lying in a finite
range of values $z_{i,c}^{min}$ and $z_{i,c}^{max}$ from a known
probability distribution $Prob_i(z_{ic} = r)$. This distribution functions
can be different at different sites. Note that while the threshold is
randomly reset, we are assuming that the matrix $\Delta$ specifying how
particles are transferred under toppling does not change.  Without any
loss of generality, we may assume that $z_{i,c}^{min} = \Delta_{i,i}$.  
This corresponds to setting the minimum allowed value of $z_i$ to be zero
for all sites $i$.

It is easy to see that this model is still abelian, and that just like the
ASM with deterministic rules, there are forbidden subconfigurations in 
this
model. In fact the forbidden subconfigurations are exactly the same as in
the deterministic ASM specified by the toppling martix $\Delta$
\cite{dd90}. The number of allowed configurations in the stochastic model
is larger, as the maximum allowed value of $z_i$ is larger. We define a
new matrix $\tilde{\Delta}$ such that
\begin{equation}
\tilde{\Delta}_{i,j}=\Delta_{i,j}, ~~for~~ i~ \neq j;
\end{equation}
and
\begin{equation}
\tilde{\Delta}_{i,i}= z_{i,c}^{max}.
\end{equation}
Then, it is easy to see that any allowed configuration for a deterministic
ASM with toppling matrix $\tilde{\Delta}$ is also a recurrent
configuration of the stochastic model, and vice versa. The number of
recurrent configurations is given by $det( \tilde{\Delta})$.  However, all
configurations are not equally likely to occur in the steady state.
 
For the 1-dimensional model studied in this paper, the only non-zero
entries of $\tilde{\Delta}$ are $\tilde{\Delta}_{i,i}=3$, for $i < L$, and
$\tilde{\Delta}_{L,L} = 2$, and $\tilde{\Delta}_{i,i\pm 1}= -1$. The
determinant is easy to calculate, recovering the results of \cite{oslo1}.

Again, for this stochastic model, we define operators $\{a_i\}$
corresponding to adding a particle at site $i$, and relaxing. Then these
operators commute with each other. In addition, they satisfy the equations
\begin{equation}
\label{red'}
a_i^{z_{i,c}^{max}} = a_i^{(z_{i,c}^{max} - z_{i,c}^{min})} \prod_{j \neq 
i} 
a_j^{- \Delta_{i,j}}.
\end{equation}

This equation is a generalization of the Eq.(\ref{reduce}) in the previous
section. As before, we find that the eigenvalues of the operators
$\{a_i\}$ are either all zero, or same as that for the deterministic ASM
with toppling matrix $\Delta$. The steady state $|\Psi_{st}\rangle$ of the
stochastic model is given by
\begin{equation}
|\Psi_{st}\rangle = \prod_i  a_i^{(z_{i,c}^{max} - z_{i,c}^{min})} 
|\Phi_{st}\rangle,
\end{equation}
where $|\Phi_{st}\rangle$ is the steady state of the deterministic ASM with
toppling matrix $\Delta$. To prove this statement, we note that if for the
deterministic model $a_i|{\cal C}\rangle = |{\cal C'}\rangle$, then for the stochastic
model we must have
\begin{equation}
a_i \prod_i  a_i^{(z_{i,c}^{max} - z_{i,c}^{min})} |{\cal C}\rangle = \prod_i  
a_i^{(z_{i,c}^{max} - z_{i,c}^{min})} |{\cal C'}\rangle,
\end{equation}
as each toppling which is required to relax from ${\cal C}$ to ${\cal C'}$
in the deterministic case, will also be allowed in the stochastic case
with the larger threshold values, so long as there are enough extra
untoppled particles present\cite{footnote}. Then, if $|\Phi_{st}\rangle$ is an
eigenvector of all $a_i$ with eigenvalue $1$ in the deterministic model,
$|\Psi_{st}\rangle$ would be so  for the stochastic one.

It may be noted that this operator algebra does not depend on the
probability distribution of the random thresholds, and depends only on the
values of $z_{i,c}^{max}$ and $z_{i,c}^{min}$, and the toppling matrix
$\Delta$.  All nontrivial eigenvalues of $\{a_i\}$ are solutions of
algebraic equations (\ref{red'}), and are the same as for the
corresponding ASM with toppling matrix $\Delta$.

The Oslo rice-pile model is special in that in this case, the
deterministic ASM with same matrix $\Delta$ has only one recurrent
configuration, and the $|\Phi_{st}\rangle = |1,1,1,\ldots\rangle$. This explains why
${\cal W}$ has only one non-trivial eigenvalue.

The distribution of avalanche sizes, and the weights of different
configurations in the steady state certainly do depend on the probability
distribution of thresholds. These are rather difficult to calculate
explicitly even for the original one-dimensional model, except for very
small values of $L$.  Exact calculation of these for general $L$ seems
more difficult. This seems to be an interesting open problem.


\begin{thebibliography}{999}
\bibitem{footnote} We cannot cancel the products over $i$ on both sides
of the equation, as inverses of the operators $a_i$ are not defined.
\end{thebibliography}
\end{document}